\documentclass[prb,preprint]{revtex4-1} 
\usepackage{amsmath}  % needed for \tfrac, \bmatrix, etc.
\usepackage{amsfonts} % needed for bold Greek, Fraktur, and blackboard bold
\usepackage{graphicx} % needed for figures
\usepackage{lineno}
\begin{document}

% Be sure to use the \title, \author, \affiliation, and \abstract macros
% to format your title page.  Don't use lower-level macros to  manually
% adjust the fonts and centering.

\title{Measuring Sound, One Ping at a Time }
% In a long title you can use \\ to force a line break at a certain location.

%When submitting the manuscript for review, do not include the author's name or institution
\author{Helio Takai}
\email{htakai@pratt.edu} % optional
\altaffiliation[Pratt Institute ]{200 Willoughby Ave, Brooklyn, USA}
\affiliation{Department of Mathematics and Science, Pratt Institute, Brooklyn, NY 11205}

\author{Tom Tomaszewski}
\affiliation{Shoreham Wading River HS, 250B NY-25A, Shoreham, NY 11786 (retired)}

\author{Jeremy Tomaszewski}
\affiliation{Haverford High School, 200 Mill Rd, Havertown, PA 19083}

\author{Joe Sundermier}
\affiliation{Deer Park High School, 1 Falcon Pl, Deer Park, NY 11729 (retired)}
 % optional second address
% If there were a second author at the same address, we would put another 
% \author{} statement here.  Don't combine multiple authors in a single
% \author statement.
%\affiliation{Department of Physics, Weber State University, Ogden, UT 84408-2508}
% Please provide a full mailing address here.

%\author{David P. Jackson}
%\email{ajp@dickinson.edu}
%\affiliation{Department of Physics, Dickinson College, Carlisle, PA 17013}

% See the REVTeX documentation for more examples of author and affiliation lists.

 \date{\today}

\begin{abstract}
Understanding how sound propagates through different media is fundamental to both science and technology. While sound plays a critical role in natural navigation and underlies a wide range of modern applications—from medical ultrasound to sonar and gas analysis—its teaching in classrooms often remains limited to traditional and abstract demonstrations. This paper presents an accessible, low-cost experimental setup that enables students to measure the speed of sound in gases using time-of-flight techniques. Constructed from common materials and powered by open-source software, the device provides accurate, hands-on measurements while allowing students to explore the effects of temperature, pressure, and gas composition. By making acoustic wave physics tangible and interactive, this approach fosters deeper conceptual understanding and promotes active experimentation in diverse educational environments.
\end{abstract}
% AJP requires an abstract for all regular article submissions.
% Abstracts are optional for submissions to the "Notes and Discussions" section.

\maketitle % title page is now complete

\section{Introduction} % Section titles are automatically converted to all-caps.
% Section numbering is automatic.

Sound is a mechanical wave that travels through gases, liquids, and solids, and its speed depends on the medium’s temperature, pressure, and composition. In nature, animals such as bats and dolphins rely on echoes to navigate with striking precision. Humanity’s grasp of sound propagation now underpins technologies from medical ultrasound and industrial cleaning to geological mapping, sonar, gas‐purity monitoring, and even assistive devices that help people with visual impairments move confidently through busy streets. Acoustic ranging has become so commonplace that anglers locate schools of fish with sonars\cite{fishsonar}, and pop-culture heroes like Daredevil use fictional echo-vision to “see” their surroundings\cite{daredevil}.

In many classrooms, acoustics is still presented with the same demonstrations that teachers used decades ago—most often the resonant tube\cite{resonant} or the Kundt tube\cite{kundt}. Valuable as these classics are, they rarely give students an intuitive feel for how sound actually travels or how its speed changes with environmental conditions. In other words, how deeply sound connect with the media it propagates. Precise speed-of-sound measurements are critical for understanding mechanical waves, but students seldom get the chance to make them.

This paper introduces a easy to assemble setup that lets learners measure the speed of sound in gases accurately and hands-on. Built from affordable, readily available materials and driven by open-source software, the apparatus works on a lab bench, a classroom desk, or even a kitchen table. Students can record time-of-flight data and experiment with sound propagation in different gases.  In the process they move beyond abstract formulas, testing wave theory with real data they have gathered themselves.

By putting good accuracy within reach of a classroom budget, this experiment turns acoustics into an active inquiry. It invites learners to explore how sound navigates the world—one ping at a time—and makes wave physics both tangible and engaging across a wide range of educational settings. 

\section{Experimental Setup}

The time-of-flight (TOF) approach offers students a direct, intuitive way to measure the speed of sound. A short acoustic pulse travels through air and reaches two locations at different times; dividing the known microphone spacing by this time delay yields the propagation speed. Classroom implementations follow two main strategies: (i) two-microphone systems that record the pulse at separate points along its path \cite{tof1,tof2,tof3,arbor}and (ii) single-microphone systems that capture the initial pulse and its echo from a reflective surface\cite{echo1}. Both configurations achieve excellent precision when their waveforms are displayed on a computer. Tobias da Silva and Aguiar, implements a single microphone and a long garden hose to record the outgoing pulse and its delayed exit signal, giving students a clear, quantitative sense of how sound moves through air\cite{hose1}. They have surveyed the students who participated in the experiments to find that the experiment helped their understanding of the physics of sound propagation. 

\begin{figure}[h!]
\centering
\includegraphics[width=5in]{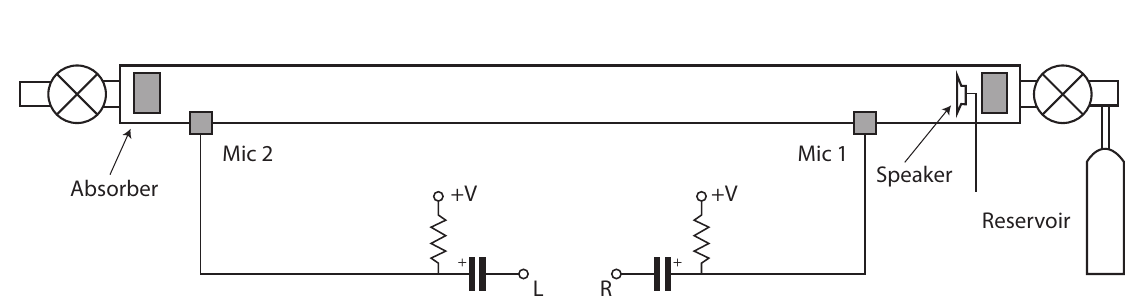}
\caption{Diagram for the apparatus for the measurement of the speed of sound in gases. }
\label{diagram}
\end{figure}

Building on that pedagogical theme, this paper presents a low-cost, high-precision time-of-flight (TOF) apparatus for measuring the speed of sound in gases. The core of the instrument is a rigid, 2-inch-diameter PVC tube sealed with end caps to form a closed test chamber. A 1-inch speaker mounted at one end generates repeatable acoustic pulses (“pings”), while two 5 mm electret microphones are epoxied into indentations along the wall, each aligned with a 2 mm hole into the tube’s interior and spaced $200 \pm 1\text{mm}$ apart. Mounting the microphones externally helps preserve waveform integrity, avoiding the distortion seen in early prototypes. A schematic of the setup is shown in Figure~\ref{diagram}. Standard valves on each end cap allow the chamber to be flushed with different gases. To suppress echoes in this sealed environment, crumpled tissue paper is placed at both ends as a simple but effective absorber. Smooth-walled cardboard shipping tubes also perform well, provided the interior is clean and unobstructed to maintain signal clarity.

For the measurement, a "ping" is produced by the speaker at $1~\rm{s}$ interval. The "ping" is generated in 
 Audacity\cite{audacity}(an open source audio recorder and editor) tone-synthesis mode: a 10 cycle, 2\;\text{kHz} sine wave shaped with an asymmetrical Gaussian envelope and repeated once per second. The file is saved as \texttt{.wav} or \texttt{.mp3} and played back with any standard media player. This waveform has a 17~cm wavelength which is adequate for this tube length. There was no attempt to optimize the signal frequency for the measurement. 
 
\begin{figure}[h!]
\centering
\includegraphics[width=3in]{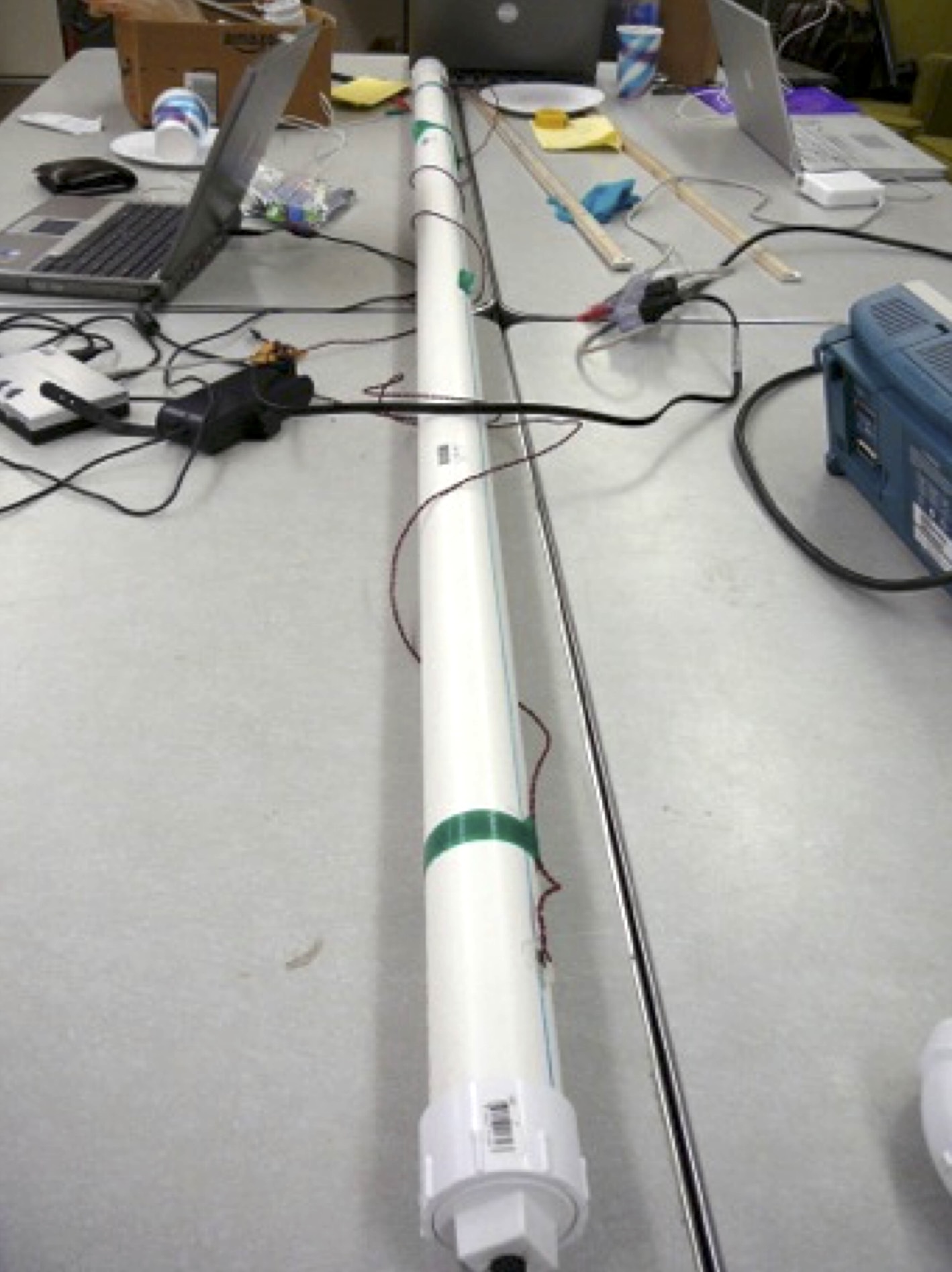}
\caption{Image of the apparatus to measure the speed of sound in gases. The tube measures a bit over 2 meters,
with microphones placed precisely at 2 meters. }
\label{tube}
\end{figure}

Waveforms are captured with Sound Card Oscilloscope ( by C. Zeitnitz ) that is distributed free of charge for educational institutions\cite{zeitnitz}. It emulates a digital oscilloscope using the sound input as the source of waveforms. Although this is the choice for our measurements, Audacity  will give the same results. The electret microphones are wired in stereo, powered by a 9\;\text{V} battery with a $2.2\;\text{k}\Omega$ bias resistor and a $10\;\mu\text{F}$ coupling capacitor (Fig.~\ref{diagram}). Because many computers now lack stereo line-in jacks, an external USB sound card is recommended. Generally, only one measurement is made for each setup - "one ping only". However, with Audacity it would be possible to record many waves and multiple measurements could be done. When using Audacity, instead of measuring time, it is preferable to measure the number of samples and use the sampling frequency to convert to time interval. 

\begin{figure}[h!]
\centering
\includegraphics[width=4in]{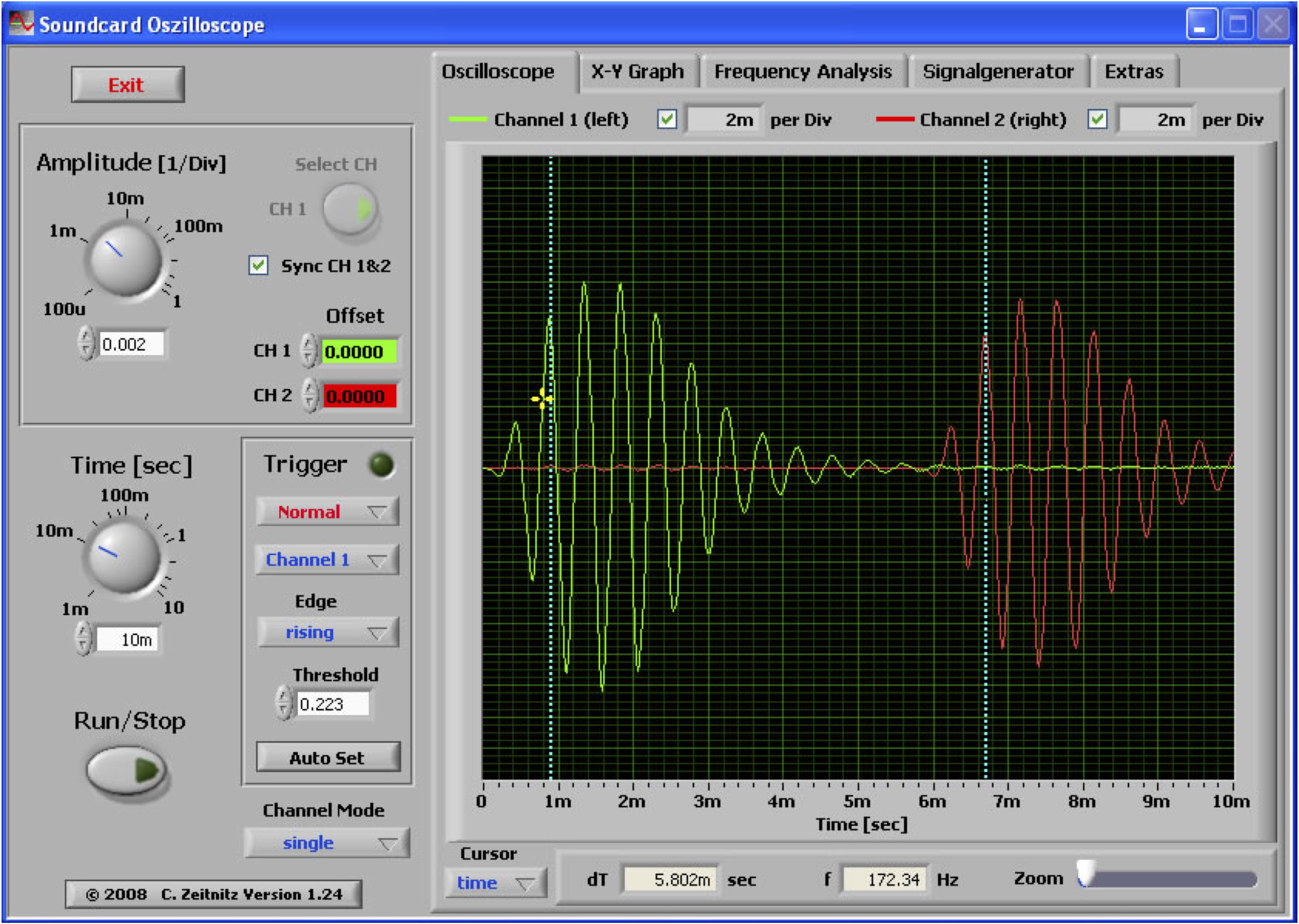}
\caption{Waveforms for the two microphones in the tube for the measurement of speed of sound in air. The green trace is for the microphone placed near the speaker and the red for the microphone place at 200 mm distance.}
\label{waveforms}
\end{figure}

Figure~\ref{waveforms} shows typical traces: the green pulse arrives at microphone 1 (nearest the speaker) before the red pulse reaches microphone 2 (further from the speaker). As the wave propagates linearly in the tube there is no observed phase alteration. Some amplitude will be lost during the propagation. For the distance used the loss is not a concern. There is no observed spread in frequency, which allows for a precise measurement of time difference by measuring the distance between two local wave maxima. 

\section{Experimental Data and Results}

Using this apparatus, the speed of sound was measured in four gases: air, $\mathrm{CO_2}$ (from a paintball cylinder), high-purity helium (laboratory grade), and commercial balloon helium. Signal amplitude varied with gas density—helium produced the weakest signal, though it remained clearly measurable. Of the gases tested, high-purity helium is the least accessible in school settings, while $\mathrm{CO_2}$ is easily sourced from soda-stream cartridges, electronics dusters, or paintball cylinders.

Table~\ref{speedgasses} summarizes the measured values alongside those predicted by the ideal gas law and theoretical calculations. All measurements were taken at a room temperature of $21.8^{\circ}\text{C}$. The results show excellent agreement with expected values and clearly demonstrate how the speed of sound varies with each gas. This offers students a hands-on and intuitive way to explore the physics of acoustic propagation with high precision.

\begin{table}[h!]
\centering
\caption{Speed of Sound in Gases ( @$21.8^{\circ}\rm{C}$)}
\begin{ruledtabular}
\begin{tabular}{l c c c c c }
Gas & Density & $\gamma$ & Measured & Ideal Gas & Empirical\\
        & ($\rm{g/cm}^3$) &      ($\rm{C}_P/ \rm{C}_V$)       & (m/s) &    (m/s)   & (m/s) \\
\hline	% horizontal line to separate headings from data
Air         &  $1.292\times 10^{-3}$    &  1.400    &  344.7 & 344.3 & 344.2 \\
$\rm{CO}_2$ &  $1.977\times  10^{-3}$ &  1.297 & 267.0 & 267.9 & 267.0 \\
Helium & $0.179 \times 10^{-3}$ & 1.667 &  1012.7 & 1011.5  & 1010.1 \\
Balloon Helium &   ---  & ---   & 837.9 & ---  & ---   \\
\end{tabular}
\end{ruledtabular}
\label{speedgasses}
\end{table}

The speed of sound in an ideal gas is given by:

\begin{equation}
v (m/s) = \sqrt{ \frac{\gamma R T}{M}},
\label{idealgas}
\end{equation}

where $\gamma$ is the adiabatic index ($C_p/C_v$), $R$ is the ideal gas constant (8.314 J/(mol·K)), $M$ is the molar mass in kg/mol, and $T$ is the absolute temperature in Kelvin.

The experimental values have small uncertainties and agree well with both empirical and ideal gas predictions. 
The empirical values are commonly found scaling the speed of sound at $0^{\circ}$C to the temperatures of interest with the simplest scaling being by the ratio of temperatures.  For air, the agreement is within 0.1\%; for helium, within 0.25\%. They represent  good results for an apparatus to be used in the educational setting. 

The measurement also reveals that commercial balloon helium is likely not pure. Its significantly lower speed of sound suggests a mixture with other heavier gas, most likely air. For gas mixtures, the effective speed of sound can be estimated by calculating the equivalent molar mass $M_{eq}$ and adiabatic index $\gamma_{eq}$ based on the fractional pressure contributions \cite{gasmix}:

\begin{equation}
\gamma_{eq} = \frac{(1- \epsilon)C_{p1} + \epsilon C_{p2}} { (1- \epsilon)C_{v1} + \epsilon C_{v2} },
\label{CpCveq}
\end{equation}

and,

\begin{equation}
M_{eq} = (1-\epsilon)M_1 + \epsilon M_2,
\label{meq}
\end{equation}

where $\epsilon$ is the ratio of the partial pressures of Helium and Air.  The specific heat capacities in J/(kg·K) are: $C_p = 1.005$, $C_v = 0.718$ for air, and $C_p = 5.1926$, $C_v = 3.1156$ for helium. Assuming air as the contaminant, a value of $\epsilon = 0.9275$ explains the measured sound speed in the balloon helium sample. This estimate aligns with reported purities of commercial helium used for inflating party balloons. However, this ratio will depend on the global helium availability and it could be lower\cite{balloon}. 

\section{Discussions and Conclusions}

The apparatus presented here offers an accessible, low-cost solution for measuring the speed of sound in gases. Constructed from readily available components and paired with open-source software, it serves as a versatile platform for hands-on experimentation. Despite its simplicity, the setup achieves a high level of precision, with results that closely align with both empirical data and theoretical predictions based on the ideal gas model at room temperature. This accuracy reinforces key physical principles—particularly the dependence of sound speed on temperature. The ability to distinguish between pure and commercial-grade helium, with the latter exhibiting a slower propagation speed despite retaining buoyancy, highlights the system’s sensitivity and educational value.

Beyond basic measurements, the device supports more advanced analytical extensions. For instance, phase analysis between the two recorded waveforms can be implemented to determine the time delay with greater accuracy or to examine subtle waveform spreading that may not be visible to the eye. These types of investigations are especially suitable for college-level coursework and offer an introduction to signal processing techniques in physics.

The instrument also lends itself to broader experimental inquiries. Incorporating a thermometer would enable systematic studies of temperature dependence, while introducing gas mixtures in known proportions allows exploration of how molecular composition affects acoustic propagation. Such extensions promote deeper discussions about the relationship between wave behavior and material thermal and pressure properties.

In summary, this compact, affordable, and adaptable instrument provides a powerful educational tool for exploring acoustic physics. Its modular design and intuitive data visualization encourage students to move beyond passive learning, engage directly with core physical concepts, and develop a deeper appreciation for the dynamics of wave propagation.

\begin{acknowledgments}

This work was partially developed as a course project in PHY-579 (Special Topics) offered by the Department of Physics and Astronomy at Stony Brook University, and further advanced during the QuarkNet Summer Program and Summer Undergraduate Laboratory Internship (SULI) programs at Brookhaven National Laboratory. 

\end{acknowledgments}

\end{document}